\documentclass[twocolumn,aps,prl,showpacs,superscriptaddress]{revtex4}

\begin{document}

\title{Comments to topological defects in bilayer vesicles}

\author{L. V. Elnikova}

\address{
A. I.~Alikhanov Institute for Theoretical and Experimental Physics, \\
25, B.~Cheremushkinskaya st., 117218 Moscow, Russia}

\date{\today}

\begin{abstract}
To explain the details of bilayer vesicle aggregation, we revised
the anyon model for lipid domains formation in closed vesicles of
lipid-cholesterol system $DPPC/DLPC$/cholesterol, which was
measured by Feigenson and Tokumasu (Biophys. Journal, 2001, 2003)
in frames of the different optical experiments and atomic-force
microscopy.
\end{abstract}


\maketitle

\textbf{Introduction}

The idea by Park with coworkers \cite{EPL1992}, and by Evans
\cite{Evans95} about the orientational order parameter in closed
liquid crystal aggregates of spherical and non-spherical shapes,
has been almost illustrated on the Feigenson's experiments
\cite{Feig}. The nanoscopic domains formation of the giant
unilamellar vesicles (GUV's) surface, consisting of
dipalmitoylphosphatidylcholine (DPPC)/dilauroylphosphatidylcholine
(DLPC)/ cholesterol mixed bilayers, were explained by the
fractional quantum Hall effect analogy \cite{EL0601651}.

However, the confocal fluorescent microscopy (CFM) and the
fluorescence resonant energy transfer (FRET) measurements
\cite{Feig} are not capable to identify the real sizes of these
surface structures, predicted for the "D" - region of the phase
diagram \cite{Feig}. Using the atomic force microscopy (AFM) on
the large unilamellar vesicles (LUV's) \cite{Feig2003}, the
regions of nanoscopic domains, as well as of microscopic domains,
became observable in "A" , "B" , and "C"  - regions in addition,
where the inter-layer bulk lipidic clusters were found. Applied
LUV suspension on freshly cleaved mica \cite{Feig2003} in AFM
measurements render a native GUV's environment of the same lipids
closely resembling.

The data of \cite{Feig2003} revealed the significant features of
the domains formation. If inside of the "A" , "B" , and "C"
bilayer regions, the lipidic clusters exist \cite{Feig2003}, the
model with anyons, in which FQHE formalism has been used
\cite{EL0601651}, may be corrected.

Now looking to \cite{Feig2003}, disordered closed vesicle axes
stabilization is caused not only by spontaneous appearance of the
surface lipid domains, but also by internal bilayer forces,
forming the bulk defects (hedgehogs).
 Hedgehogs prevent us to use the bilayer thickness as a parameter,
 which is connected the manifold of two spheres $M_1 = S_1^2$ and
$M_2=S_2^2$, corresponding to an outer- and an inner monolayer of
a closed vesicle, respectively.

We suspect the composition of all observed defects can be
described in frame of non-Abelian statistics (that could be
applied only for surface vortexes alone). These common topological
properties are observable in superfluid $He$ phases, liquid
crystals and superconductors \cite{Evans95, MM, Volovik, Kurik}.

Since the geometry of the inner GUVs sphere is not viewed by means
of CFM et. c. \cite{Feig}, and only the nanoscopic clusters of
about 46 nm diameters are dispersed throughout of all areas in
LUVs, we can discuss the different mechanisms of fractional axis
shape stabilization, associated with defects of both types.

\textbf{1. Formalism}

Using mean-field approach for lowest Landau levels with some gauge
parametrization, Evans found \cite{Evans95} the partition function
of the vesicle
\begin{equation}
Z=2\pi\int e^{-F_{eff}(t,\mathbf{n_1},\mathbf{n_2})}t^5 \\ (1-
\mathbf{n_1}\mathbf{n_2})dtd\Omega_1d\Omega_2\delta(\mathbf{n_1}\mathbf{n_2}-\chi)d\chi,
\end{equation}
where $F_{eff}=F_{eff}(\chi,t)$ is the effective potential,
$-1\leq\lambda\leq1$ is a measure of the relative separation of
two vortexes, $d\Omega_k=sin\theta_kd\theta_kd\varphi_k$, with the
orientational variables in spherical coordinates $\theta_k$,
$\varphi_k$. 'Modified normal' unit vectors from the center of
vesicle toward the $k^{th}$ vortex obey
$\mathbf{n_1}\mathbf{n_2}=\cos\theta_1\cos\theta_2+\sin\theta_1\sin\theta_2\cos(\varphi_1-\varphi_2)$,
$t$ is a measure of the overall real amplitude of the order
parameter.

(The basic parametrization is given by the measure
\begin{equation}
\textit{D}[\rho]=\prod_\Omega \mathbf{\hat{r}} \cdot \mathbf{N}
d\rho(\mathbf{\sigma}),
\end{equation}
where $\Omega$ are all solid angles, $\rho(\mathbf{\sigma})$ is
the scalar field of two-dimensional surface coordinates
$\mathbf{\sigma}=(\sigma_1,\sigma_2)$, and $\mathbf{\hat{r}} \cdot
\mathbf{N}=1-\frac{1}{2}\{ (\partial_{\theta}\rho)^2
+\csc^2\theta(\partial_{\varphi}\rho)^2\}$.)

All this means the surface vortexes generate the set of six
thermodynamical parameters, which can be presented in three more
or less physically combinations (of 'temperature', 'malleability',
and the scalar parameter) \cite{Evans95}. And therefore the
 filling factor $\nu$ \cite{Haldane1}
takes the easy equatable values in the anyons terms.

\textbf{2. Several ways of vesicle shape evolution under
hedgehogs}

In bulk $Sm-C$ aggregates, hedgehogs having the group
$\pi_2(S_2/Z_2)$, may be of hyperbolic and spherical structures
and be described in torus degeneracy space \cite{Oshikawa, Ei}.
They always can be taken up (let out) by the vesicle surface
vortexes \cite{Garel, Trebin}.

So with the advent of the hedgehogs, at certain thermodynamical
conditions (temperature and/or lyotropic content of a vesicle), it
is possible to find the next positions of defects, and of the
anyon states.

\textbf{2.1. Coexisting of hedgehogs with surface vortexes}

Here all hedgehogs play role in an external potential for Laughlin
wave function \cite{Haldane1}.

\textbf{2.2. Destruction of hedgehogs}

In case of rasing of the hedgehogs to the (for instance, to the
outer $S^2_1$) surface, the effect of influence $\pi_1(R)$ group
onto $\pi_2(R)$ can be observed \cite{Kurik}, since $\pi_1(R)$
group is non-trivial. According to \cite{Ostlund}, the field lines
between hedgehogs can collapsed to 'string', connecting a hedgehog
with a surface vortex.

\textbf{Comments}

At continuous phase transition in the vesicle, caused by variation
of aggregate content, the broken symmetry can be quite observable
(at $\pi_2 \longrightarrow\pi_1$ homomorphism), in agree with
marginal type-I-type-II vesicle \cite{Evans95}.

As fractionalization associate with the groundstate degeneracy
\cite{Oshikawa}, the scenario in the hedgehogs presence is
noncontradictory against a background of anyons.

\textit{Acknowledgements}

The author thanks to F. Tokumasu for explaining of new phase
diagram on LUVs, and to the anonymous referee for pointing data of
\cite{Feig2003}.


\begin{thebibliography}{999}

\bibitem{EPL1992} J. Park, T. C. Lubensky, and
F. C. MacKintosh, Europhys. Lett. {\bf20(3)}, 279 (1992).
\bibitem{Evans95}R. M. L. Evans, Phys. Rev. E. {\bf53}, 935 (1996).
\bibitem{Feig} G. V. Feigenson and J. T. Buboltz, Biophys. Journal {\bf80}, 2775 (2001).
\bibitem{EL0601651}L. V. Elnikova, E-print archives, cond-mat/0601651.
\bibitem{Feig2003} F. Tokumasu, A. J. Jin, G. V. Feigenson,
and J. A.~Dvorak, Biophys. Journal {\bf84}, 2609 (2003).
\bibitem{MM} M. Monastyrsky, \textit{Topology of Gauge
Fields and Condensed Matter}, Springer (1993).
\bibitem{Volovik} G. Å. Volovik and V. P. Mineyev, Zh. Eksp. Teor. Fiz. {\bf72}, 2256 (1977) (in
Russian).
\bibitem{Kurik} M. B. Kurik and O. D. Lavrentovich, Uspekhi Fiz. Nauk {\bf154}, 381 
(1988) (in Russian).
\bibitem{Haldane1}F. D. M. Haldane, Phys. Rev.
Lett. {\bf51}, 605 (1983).
\bibitem{Oshikawa}M. Oshikawa and T. Senthil, Phys. Rev. Lett. {\bf96}, 060601 (2006).
\bibitem{Ei}T. Einarsson, Phys. Rev. Lett. {\bf64}, 1995 (1990).
\bibitem{Garel} A. T. Garel, J. de Phys. {\bf39}, 225 (1978).
\bibitem{Trebin} H.-R. Trebin, Adv. Phys. {\bf31}, 195 (1982).
\bibitem{Ostlund} S. Ostlund, Phys. Rev. B. {\bf24}, 485 (1981).

\end{thebibliography}
\end{document}